# Wireless Sensor Networks: Local Multicast Study

Seyed Hossein Ahmadpanah

Department of Computer and Information Technology,
Mahdishahr Branch, Islamic Azad University, Mahdishahr,
Iran

Abdullah Jafari Chashmi

Department of Electrical and Telecommunications
Engineering Technology, Mahdishahr Branch, Islamic Azad
University, Mahdishahr, Iran

Seyede Samaneh Siadatpour

Department of Computer and Information Technology,
Mahdishahr Branch, Islamic Azad University, Mahdishahr, Iran

*Abstract*—Wireless sensor networks and Ad-hoc network in the region Multicast (Geocasting) means to deliver the message to all nodes in a given geographical area from the source point. Regional Multicast practical application of the specified area and other regions may be formed to broadcast transmission and location-related business information, extensive advertising, or to send an urgent message. Regional multicast protocol design goal is to ensure messaging and low transport costs. Most have been proposed agreement does not guarantee message delivery, although some other guaranteed messaging protocol has triggered high transmission costs. The goal now is to ensure that research messaging and low-cost local transmission protocol and its algorithm, to promote the development of domain communication. This paper introduces the research background and research results, and proposed to solve the problems and ideas.

I. INTRODUCTION

Wireless sensor network is Business Week and MIT Technology Review as one of the 21 most influential technology and change the world's 10 largest technology of the 21st century, the future is an important part of the network. It can be used for environmental monitoring, incident tracking, safety and control systems it can also be considered as a distributed database system that supports different types of queries. One type is called requirements are based on local queries all sensors within a geographic area to participate, which will help to build a query response. For example, queries a specific area within the inner position and then two hours for all vehicles. To support such a query, the monitoring center (receiver) should be such a query message to all sensors query geographical. The source node sends a message to a given geographical area all the way nodes is called the local Multicast (Geocasting) [1]. Ad hoc networks and sensor networks, local broadcast group are many network applications in basic services.

Multicast region has many interesting applications. It can form a regional broadcasting transmitting and location-related business information, advertising, etc., in order to reach such a vehicle in a geographical area to send parking location within the scope of free parking and other information. Also send an emergency message to a specified area is also one of its important applications.

Regional multicast algorithm target consists of two parts: guaranteed message delivery and low transport costs. Guaranteed message delivery to ensure you can receive a copy of the local multicast message in the query Territory each sensor; low cost of transmission can save energy as much as possible, and cannot adapt to the energy limited in most cases to replace the battery sensor needs. In order to achieve geographical broadcast group, many algorithms [1-10] has been proposed researchers. But some methods [2-9] proposed does not guarantee delivery of messages, and cause higher transport costs. [1] and four algorithms [10] proposed guaranteed message delivery, but high transport costs. In all regions multicast algorithm, the dividing plane crossing limiting flooding method RFIFT (Restricted Flooding with Intersected Face Traversal) [1] is the most effective one.

II. CURRENT RESEARCH AND ANALYSIS

*A. Background and History*

Multicast geographical concept was first in 1997, Navas and Imielinski [12] suggested that the fundamental problem is that the message is sent from a source to a certain geographic area (such as a building, a street, a school, a business area All nodes, resorts, etc.), also known as a broadcast location-based [13]. Because the goal of many communication applications is a given reach all nodes in the region, so the region is Multicast Ad hoc networks and wireless sensor network is an important means of communication, for example, such as contact sensors monitor all active areas to regularly collect data, or for reporting purposes will inform sensors cover a given area of its location.

The first work in the MANET Geocasting in the literature [2] proposed, the main idea is to flood the pan constraint-based forwarding. Location information source host and destination domain determines the forwarding flooded areas. Typically forwarding area is defined as the smallest rectangle including the sender and destination domain location. When forwarding node regions received Geocast message will be re-broadcast the message, not in the area when the node forwards messages received Geocast discard the message. But in this way there are still unnecessary flooding packet forwarding area.

Many local multicast protocols have been proposed [3-9,13,14], but cannot guarantee delivery to all nodes multicast geographical area within R. Dispersed within a geographic area of the sensor may be due to differences in perception and communication barriers or radius region do not cause them to be connected to each other. One kind " 'forgotten' face tree traversal scheme)" [11,15] can ensure the transmission of information, but the information has a considerable overhead.

(2) has been proposed geographic multicast routing protocols have been proposed to classify geographic multicast routing protocol [16] will be divided into three categories:

A- protocol based on flooding (flooding-based protocols): Use flooding or flooding will change the geographical multicast packets forwarded from the source to the local multicast area, limiting flooding forwarding area including source and geographical multicast the minimum rectangular area. Such agreements include Location-Based Multicast (LBM) [2], Voronoi diagram based geocasting protocol [5], Flooding-Based Geocasting Protocols for Mobile Ad Hoc Networks [6]. These agreements even reduce the flood area, but still cause high cost of flooding, but does not guarantee message delivery [1].

B - based routing protocol (routing-based protocols):. generating a route from a source to a geographical area by controlling multicast packets. Such agreements include Mesh-based Geocast Routing Protocol (MGRP) [17], Geocast Adaptive Mesh Environment for Routing (GAMER) [9] and GeoTORA [18]. Seada et al [20] proposed a geographic routing mechanisms and regional flooding together to achieve a high transfer rate and low cost effective and practical local multicast protocol. Although these agreements reduce the number of message transmission, but does not guarantee delivery of messages [1].

C - Protocol-based cluster (cluster-based protocols):. MANET based on geographical information is divided into several unconnected, the cell area equal to the size, and select a cluster head to perform the exchange of information in each region. Such agreements include GeoGRID [13], Obstacle-Free Single-Destination Geocasting Protocol (OFSGP) and Obstacle-Free Multi-Destination Geocasting Protocol (OFMGP) [19]. These agreements despite the increase in the transfer rate of the message, but also caused a larger transfer the cost [1].

(3) Regional multicast integrated approach to overcome the shortcomings of the above-described area multicast protocols and algorithms in recent years, some of the location-based unicast, limiting flooding and …

Across the face (face traversal) mode integrated with algorithms have been proposed to ensure the delivery of messages [1,10,11]. Depth-first surface tree traversal method (DFFTT) [11] First, use GFG (Greedy-Face-Greedy) passing local multicast message to a node within R, then a cover and geographic multicast domain R intersecting plane tree all faces (face tree) is constructed, and finally by each node of the tree surface, the message will be delivered to all the nodes in R; RFIFT [10] will be the first to use GFG geographical multicast messages transmitted to the unicast routing based on location R within a node, then the formation of limiting flood in R and R to go through all the intersecting faces; multicast-based band entrance (entrance zone) area multicast method (EZMG) [1] the local multicast domain R the surround area is divided into a series of entrance with, when a multicast source to send a message to all the entrance with a node receives a message with the entrance on broadcasting the news, then all nodes within R have heard the news. These methods ensure the transmission of information, but also caused a large transport costs.

(4) local multicast technology

MANET in area multicast (Geo-multicasting) is a not much research in the field. Multicast region [8]

In a particular message will be delivered to the technical area of special user groups, it is a special kind of depends on the location of multicast technology that defines the node local multicast group within a particular region of a collection of some special group. [8] proposed architecture and protocol support local multicast service in mobile Ad-hoc wireless networks, and other restrictive environment, they have a high accuracy and the cost effectiveness of message passing. [8] The clustering technique different mobile nodes forming a cluster (cluster). In order to save resources and reduce energy consumption, to elect a cluster head in each cluster in the region as a multicast group member nodes. Upon receiving multicast information, the cluster head in the cluster in their broadcasts. Several existing clustering algorithms to determine the cluster head. A minimum ID algorithm [21], and the other is connected to the highest (degrees) algorithm [22], there is a mobile-based clustering algorithm (MBC) [8]; [23] proposed a motion-adaptive clustering way, through active unicast routing algorithm scored probability model for future network links available.

III. NEW IDEAS AND SOLVE PROBLEMS

Now the international research community has on a number of research areas of communication, including the shape and geographical area Multicast equation, but not perfect, there are still some problems. We believe the most critical problems and their solutions are the following ideas:
(1) Existing local multicast protocols and algorithms cannot meet the guaranteed message delivery and low transport costs

of these two basic objectives. Ensure messaging and low cost of local multicast transmission design two basic goals is conflicting, the two factors. Existing local multicast protocol cannot meet these two basic requirements, designed to ensure that both messaging and low-cost local multicast transmission protocol will be the key issues and difficult problems to solve.

(2) Existing local multicast protocols and algorithms are not well considered node mobility. This is not in line with the actual situation. This requires local multicast protocols and algorithms in consideration of the situation of the mobile node.

(3) When several positions in different geographical areas need to deliver the same message, there is no effective resolution mechanism. Further studies need to be considered in a state of node mobility, multicast communication problems in different geographical regions. We propose to establish and maintain a quasi-dynamic mobile environments multicast tree, multicast tree-demand computing concepts such adaptive design, low cost geographic multicast protocols and algorithms to meet the messaging different geographical regions.

(4) does not have a mechanism to confirm the existing local multicast protocols to improve the reliability of transmission. Further research will target is to develop a geographical broadcast reliable transport protocol, without reducing resource utilization, improve the reliability of messaging.

(5) Regional and local multicast applications also have great limitations. Geography and Geo existing broadcast multicast applications also only limited research content monitoring, emergency information delivery applications, the next step will be to develop different environments according to application requirements, applicable to all kinds of background and application areas of different industries

## IV. CONCLUSION

In this paper, the geographical area multicast and multicast protocols and algorithms have been proposed to do an analysis and presentation, and proposed the existing problems and achievements in further Solutions. We are hoping to initiate, allow more researchers to join the field of research, to jointly promote wireless sensor network and mobile Ad-hoc network of local and regional broadcast multicast group study, and as soon as possible with the actual application environment, accelerate our wireless network development and enhance international competitiveness.